# Single step synthesis and organization of gold colloids assisted by copolymer templates

Aurélien Sarrazin, Arthur Gontier, Alexandre Plaud, Jérémie Béal, Hélène Yockell-Lelièvre, Jean-Louis Bijeon, Jérôme Plain, Pierre-Michel Adam and Thomas Maurer *

*Univ Technol Troyes, Lab Nanotechnol & Instrumentat Opt, Inst Charles Delaunay, CNRS UMR 6279, F-10010 Troyes, France*

**Abstract.** We report here an original single-step process for synthesis and self-organization of gold colloids by simply incorporating gold salts into a solution prepared with Polystyrene (PS) - Polymethylmethacrylate (PMMA) copolymer, thiolated PS and Propylene Glycol Methyl Ether Acetate (PGMEA) as solvent. The spin-coating and annealing of this solution allows then the formation of PS domains. Depending on the polymer concentration of the as-prepared solution, there can be either one or several gold nanoparticles (NPs) per PS domains. For high concentration of gold nanoparticles in PS domains, the coupling between plasmonic nanoparticles leads to the observation of second peak in the optical extinction spectrum. Such collective effect could be relevant for the development of optical strain sensors in the next future.

## 1. Introduction

In nanotechnology, the fabrication process is driven by the targeted application and its specifications (size, shape, long-range ordering,...). In microelectronics, the *top-down approach* (Electron Beam Lithography [14], Focused-Ion-Etching [41], ...)- consisting in taking and etching a massive material down to the desired shape and size- met a great success[12] due its high reproducibility. Top-down processes nowadays allow to produce structures as small as 20nm in size. However decreasing sizes below this threshold becomes time and money consuming especially for guaranteeing well-defined edges over large areas [31]. In order to achieve this ambition, synthesis routes based on the *bottom-up* approach have been developed for about thirty years. In the field of plasmonics, colloidal synthesis [40] indeed easily provides large batch of gold [5] or silver [25] nanoparticles. For the moment, the key challenge remains the self-organization of colloids onto substrates [23, 24] especially when nanoparticles are not magnetic [18, 26, 28, 34, 43]. In the case of gold colloids, their organization onto substrates generally requires surface functionalization of either the substrate [6] or the nanoparticles themselves [22] via the use of organosilanes.

What is at stake today is therefore the emergence of routines for organization of colloids onto substrates or in thin films. Block copolymer templates are one of the most appealing routines for industrial applications due to their self-organization into microphase-separated domains with controlled shapes and sizes [10, 19]. The research topics about self-organization of metallic nanoparticles into segregated copolymer domains has started more than twenty years ago [7]. Indeed, copolymer thin films take advantage of not only the long range nanostructuration but also the large variety of patterns - from spheres to lamellar shapes - provided [20, 42].

A first strategy consists in using block copolymer film as lithography masks. For example, arrays of holes can be fabricated from PS-PMMA copolymer films by selectively degrading PMMA domains either via Reactive Ion Etching (RIE) [2, 3, 32] or UV-exposure [16, 27]. Following the same principle, another process consists in immersing the substrate into a solvent which dissolves one of the two polymers [13]. Then arrays of gold nanoparticles can be obtained either after metal evaporation and resin removal [16, 32] or after surface substrate functionalization and colloid deposition [13]. Another strategy started fifteen years ago [44] by Sita and coworkers, which also requires two steps, consists in directly incorporating previously functionalized nanoparticles into PS-PMMA copolymer solutions. Then their self-organization onto substrates can be guided via the Langmuir-Blodgett technique [22]. This leads to nanorings of gold colloids long range organized in the copolymer thin film.

For the past twenty years, more and more efforts have been devoted to the development of single-step synthesis and organization of inorganic nanoparticles in block copolymer templates. As a matter of fact, inorganic precursors can be selectively incorporated into amphiphilic block copolymers like PS-b-P2VP [29, 39]. For example, block copolymers like poly(ethylene oxide)-poly(propylene oxide) (PEO-PPO) can efficiently reduce and stabilize inorganic precursors like trihydrate tetrachloroaureate ($HAuCl_4$, $3H_2O$) [1, 36].
Here, we describe an original route for the single-step synthesis and templating of gold nanoparticles in PS-PMMA block copolymer thin films. This original routine is based on the affinity of gold for thiols and consists in mixing trihydrate tetrachloroaureate ($HAuCl_4$, $3H_2O$) with a PS-PMMA block copolymer solution and a small fraction of PS-SH homopolymer. This process leads to the fabrication of composite thin films with gold colloids driven into the PS domains. By increasing the density of gold colloids into the PS domains, collective

LSPR modes will appear due to strong plasmonic coupling and make such materials appealing for applications like light filters or strain sensors.

**2. Experimental section**

2.1. Copolymer template fabrication

The copolymer chosen in this study was PS-PMMA since both PS and PMMA exhibit interesting properties such as their close glass transition temperature $T_g$ of 100-110°C. In addition, The copolymer microphase separation has been well-extended studied and it has been shown that the order-disorder temperature $T_{ODT}$ was higher than 230°C though the chains of PMMA started decomposing at 220°C. Thus, the PS-PMMA block copolymer has to be annealed between 110°C and 210°C[2, 3]. The solvent used in this process was the Propylene Glycol Monomethyl Ether Acetate (PGMEA) (boiling point around 150°C) since it dissolves both PS and PMMA. Two PS-PMMA block copolymers, with the following molecular weights $M_w(PS)=27500/M_w(PMMA)=154000$ and $M_w(PS)=13000/M_w(PMMA)=96000$, were purchased from Polymer Source. PS homopolymer was purchased from Polymer Source whereas thiolated PS (PS-SH) was synthesized via atom Transfer Radical Polymerization (ATRP) followed by the end-functionalization of the PS chain with a thiol[11]. In order to get PS cylindrical domains into the PMMA matrix, the mass composition ratio should fit to 20% of PS [3]. PS homopolymer was thus added to achieve such composition ratio. The polymers were dissolved in PGMEA solvent to prepare 1 to 6%wt solutions and 2 to 10mg of trihydrate tetrachloroaureate ($HAuCl_4$, $3H_2O$) was added and mixed to the PGMEA solution. Then the solutions were spin-coated on glass substrates coated with an Indium Tin Oxide (ITO) layer. Eventually, the samples are annealed under $N_2$ atmosphere during at least 8 hours to both induce the copolymer microphase separation and synthesize the gold nanoparticles from the gold precursors. **Figure 1** proves that the nanocomposite thin film heating leads to the synthesis of gold nanoparticles and that when the PS homopolymer added to the PGMEA solution is thiolated (PS-SH), these gold nanoparticles are organized only into PS domains whereas when no thiolated PS is added, the gold nanoparticles are localized both in PMMA and PS domains. Therefore, the addition of thiolated PS in the PGMEA solution is fully requested to limit the organization of the gold nanoparticles to the inside of the PS domains. As evidenced by A.M. Ritcey and coworkers, the self-organization of gold colloids is driven by the ligand nature and/or by their sizes. The choice of these parameters may conduct the gold colloids to the air/PS/PMMA triple interface or inside PS domains [21] leading to the formation of either rings of Au NPs around the domains [22] or dense assemblies of Au NPs inside the domains. Here the choice of PS ligands guarantees the formation of dense Au NPs assemblies inside the PS domains [21]. Gold nanoparticles indeed exhibit a strong affinity with the thiol function [30] so that when gold precursors are reduced into gold nanoparticles during the annealing step, thiolated PS is being grafted to them[22, 33]. Moreover, **figures 1c** and **1d** put into evidence that the size of the PS domains is governed by the chain lengths of the block copolymer as shown in previous studies [2, 3].

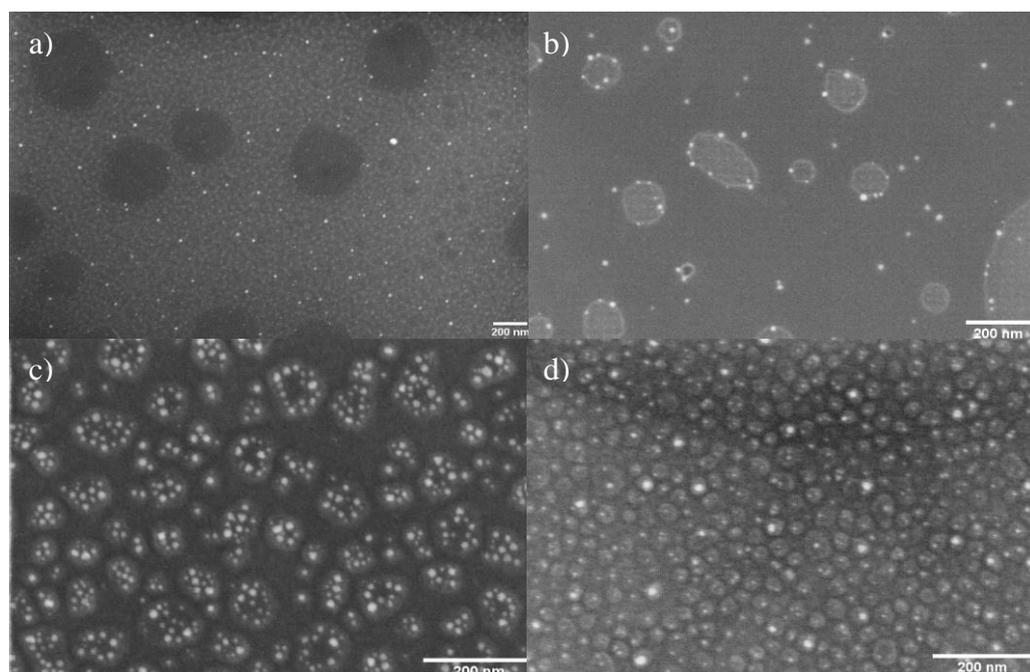

**Figure 1.** SEM images of Gold nanoparticles synthesized after microphase separation of copolymer thin films. The scale bars are 200nm large. a) no thiolated PS was added and PS domains were formed, b) no thiolated PS was added and PMMA domains were formed.

PMMA domains were obtained by using a copolymer whose PMMA chain is shorter than the PS one and by adjusting the ratio of PMMA to 20%. For a) and b), the gold nanoparticles are both in and out of the domains. c) and d) thiolated PS were respectively added to the PS(27500)-PMMA(154000) and PS(13000)-PMMA(96000) copolymers to form PS domains. The gold nanoparticles which are synthesized in-situ the copolymer thin films are templated into the PS domains.

2.2. Composite film characterization

We then studied the influence of both the polymer concentration in the PGMEA solution and the density of gold precursors added. **Figure 2** displays the aspects of the sample which were investigated. One of the advantages of such a process is the ability to fabricate either small or large substrates but also to anneal several samples in one shot. The samples are homogeneously colored and the higher the concentration of the solution and the density of gold precursors are, the stronger the coloration will be. The thickness of the composite films has been determined by AFM measurements (see **Figures 3A and 3B**). It appears that the thickness of the film increases with the polymer solution concentration from 10nm for 1%wt, 20nm for 2%wt, 65nm for 4% wt to 250nm for 6%wt (see **Figure 3C**) and polymer films become pink for at least 3%wt. The concentration of gold salts in the PGMEA has not been evidenced for playing any role in the composite film thickness though.

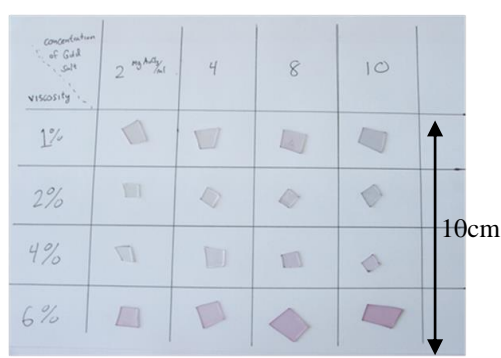

**Figure 2.** Aspect of the nanocomposite films composed of Au NPs organized into PS domains depending on both the concentration of polymers in the PGMEA solution and the density of added gold precursors. Thiolated PS was added to achieve a PS composition ratio of 20%.

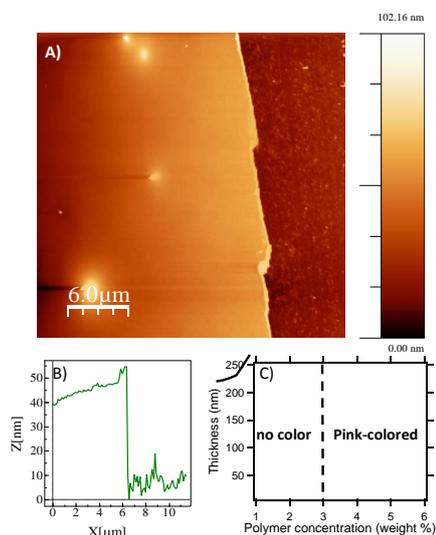

**Figure 3.** A) AFM image for polymer film thickness determination (3 %wt). B) Topography profile of a copolymer thin film obtained after spin-coating a 3 %wt polymeric solution. C) Thickness of polymer films as a function of %wt polymer solution. The polymer thin films become pink for 3 %wt or more polymer concentrations.

SEM pictures were performed on samples with different polymer solution concentrations and gold precursor concentrations (see **Figure 4**) via a FEG eLine Raith (10 kV) with an aperture opening of 10 µm and a current of 23 pA. From the SEM images analysis, information about gold nanoparticles and copolymer domains sizes may be deduced. **Table 1** presents characteristics of the composite films for two values of copolymer concentration and three values of gold salt concentration. First, the SEM images evidence the high impact of the copolymer concentration on the number of nanoparticles per PS domain. Indeed, when the copolymer concentration is above 3%, several nanoparticles per domain can be observed which makes the PS domains larger and more polydisperse (25nm for 1%wt copolymer concentration and 8mg/mL of Au salts to be compared to 50nm for 4%wt copolymer concentration and 8mg/mL Au salts concentration). This can be explained by the occupancy volume of the Au nanoparticles inside the PS domains and the variation of the Au nanoparticles per domain from one domain to the other.

As for the nanoparticle size, increasing the copolymer concentration leads to smaller nanoparticles since the mean diameter is about 20nm with a mean deviation of 5nm when the concentration is 2% or less and decreases to about 10nm with a mean deviation of 3.5nm when the concentration is more than 4% for the PS(27500)-PMMA(154000) copolymer. Thus the mean size of the nanoparticles seems to be affected by the thickness of the film, which could be explained by the possibility for Au nanoparticles to grow equally in three dimensions for films thicker than 20nm. Moreover, increasing the concentration of gold precursors does not seem to lead to larger Au NPs but to a higher average number of the latter per PS domain.

Finally, a possible explanation for the smaller size and the denser assemblies of gold colloids into domains which more strongly color the polymer films with concentration larger than 3%wt (thick films) probably lies in the greater quantity of thiolated-PS that enable more Au NPs to be trapped into PS domains and make the latter larger. Moreover the smaller nanoparticles formed in thicker films indicate that these two parameters are obviously only indirectly related through the ratio of copolymer to gold precursor: the higher the thickness of the film for a given gold salt concentration, the lower the filling fraction of the gold precursor in the film, and thus smaller nanoparticles are formed.

| | | Au salt concentration (mg/mL) | | |
|---|---|---|---|---|
| | | 2 | 4 | 8 |
| Copolymer concentration (%wt) | 1 | PS domain size=26nm<br>NP size= 18nm<br>Average of the number of NPs/domain=0-1 | PS domain size=29nm<br>NP size= 18nm<br>Average of the number of NPs/domain=1 | PS domain size=25nm<br>NP size= 20nm<br>Average of the number of NPs/domain=1-2 |
| | 4 | PS domain size=40nm<br>NP size= 8nm<br>Average of the number of NPs/domain=5 | PS domain size=51nm<br>NP size= 10nm<br>Average of the number of NPs/domain=8 | PS domain size=50nm<br>NP size= 10nm<br>Average of the number of NPs/domain=11 |

**Table 1.** SEM analysis of the composite films for PS(27500)-PMMA(154000) copolymer as a function of copolymer and Au salt concentration. The size of copolymer domains and nanoparticles as well as the number of nanoparticles per domain are provided.

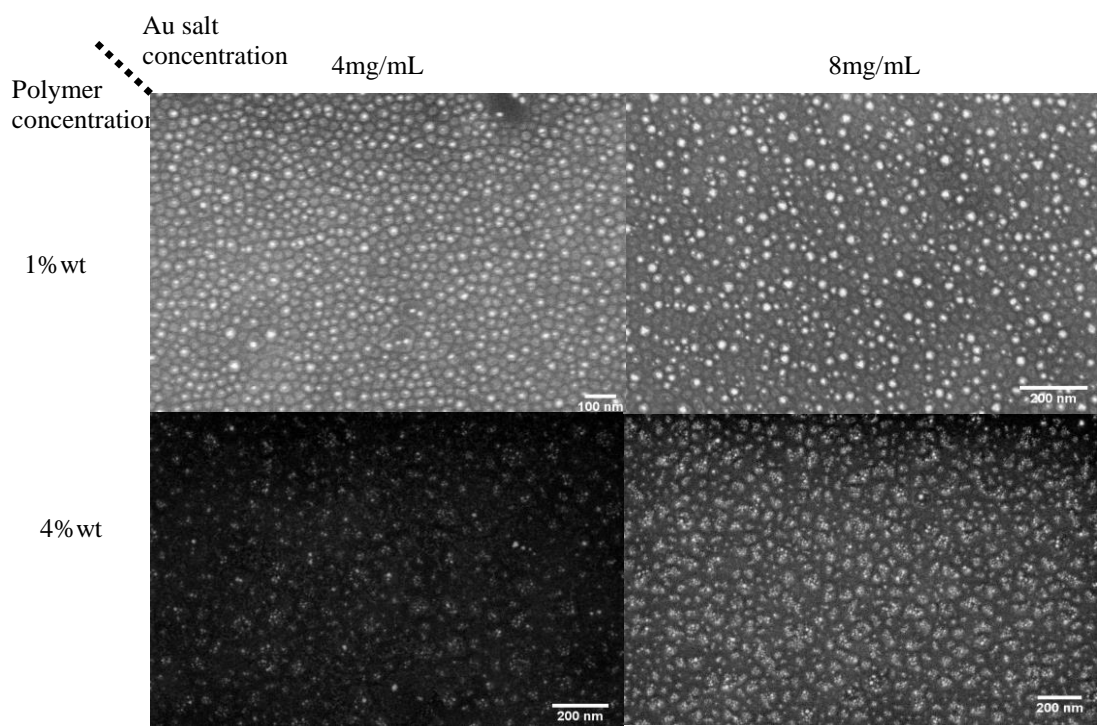

**Figure 4.** SEM images of the gold nanoparticles organized into the PS domains of the PS-PMMA block copolymer after microphase-separation (annealing at 200°C during 11 hours). The study of these images among others showed that increasing the solution concentration allows getting several Au NPs into the PS domains and that increasing the gold precursors concentration does not play any role in the Au NPs size.

## 3. Optical extinction spectra and LSPR sensors

The properties of the nanocomposite films were studied via optical extinction measurement with a transmission optical microscope coupled to a micro-spectrometer by a multimode optical fiber. Figure 4 shows the visible spectra depending on the gold precursor concentration for 4%wt copolymer solution concentration (**Figure 5a**) and 6%wt copolymer solution concentration (**Figure 5b**). These spectra all exhibit a main peak at a wavelength of about 540nm which corresponds to the predicted value by the Mie theory [4] for 10nm Au spheres in a polystyrene medium (refractive index of 1.59 in the visible range[17]). Besides, **Figure 5** also puts into evidence a peak at larger wavelength, about 740nm for 4%wt copolymer solution concentration (see **Figure 5a**) and about 700nm for 6%wt copolymer solution concentration (see **Figure 5b**). The presence of a peak at larger wavelength can be explained by longitudinal modes due to very close interacting plasmonic nanoparticles as already observed in [15]. Indeed, when nanoparticles are densely packed (dimer approximation), strong plasmon coupling can occur and gives rise to extinction peaks at larger wavelength[38]. As expected, this coupling mode is blue-shifted with increasing the concentration from 4% to 6% due to the decrease of the NP size. Increasing the gold salts concentration leads to an increase of the 540nm peak due to a higher absorption of the composite film. However, the coupling mode wavelength globally increases with increasing gold salts concentration due to a higher number of nanoparticles per domain and thus to a smaller average nanoparticle-nanoparticle distance. An exception to this tendency can be observed in **Figure 5a** for 4%wt copolymer concentration and 4mg/mL Au salt concentration. The appearance of a coupling band depends on the strength of near-field coupling between neighboring particles, which in turn depends on both distances and sizes of these same particles. Higher sizes can decrease this coupling due to increased radiative damping. A slight increase in the average coverage density of the particles can thus result in a lower coupling if particle sizes are increased. Indeed, the number of particles per domain is higher than for the composite film with 2mg/mL Au salt concentration, but the domain size too which explains why the intensity of the coupling mode may appear weak compared to the one of composite film with 2mg/mL Au salt concentration.

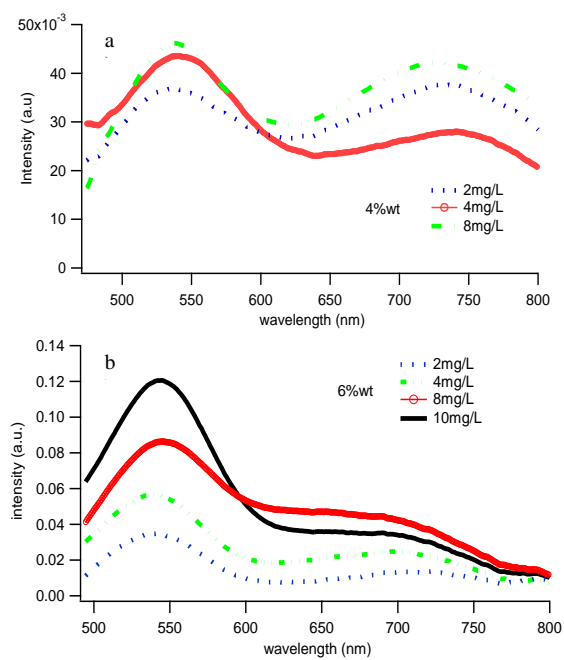

**Figure 5.** Optical extinction spectra of the nanocomposite films depending on the gold precursor concentration for a) 4%wt copolymer solution solution concentration and b) 6%wt copolymer solution concentration. On the graph b), one should notice that the 2mg/L and

4mg/L could be the most interesting concentrations since they minimize the peak of 540nm way more than the second - and more important - peak.

### 4. Numerical simulations

Discrete Dipole Approximation simulations were performed to confirm the role of interparticle coupling in the presence of the larger wavelength mode (see **figure 6**). The model which has been chosen is very simplistic but allows better understanding the apparition of the larger wavelength peak which has been rarely experimentally observed [15, 22] because requiring densely packed gold nanoparticles. Five 10nm Au NPs surrounded by PS were modeled thanks to the DDSCAT software [9] by making the distance d between two neighbouring particles vary from 0nm to 1nm. These simulations show that a second peak at larger wavelength may arise for d<1nm and that it is located between 650nm and 750nm which is experimentally observed (**see Figure 5**). The apparition of a LSPR mode at larger wavelength when Au NPs nearly touch themselves has already been theoretically studied for larger Au spheres [35] and was attributed to strong interaction between single-particle multipoles and strong oscillations of the charge pile-up near the interparticle gap. Our numerical and experimental results confirm the apparition of larger wavelength peaks due to strong coupling when Au nanoparticles are densely packed[38].

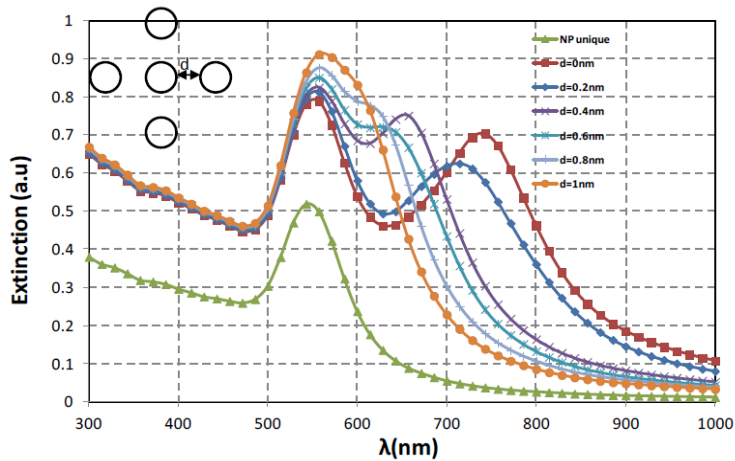

**Figure 6.** Optical extinction spectra of 10nm Au spheres calculated via the Discrete Dipole Approximation (DDA) model [9]. The optical extinction spectra are calculated for five nanoparticles displayed in staggered row and compared to the one of single nanoparticle. The surrounding medium is polystyrene and the distance d between the edges of two particles varies from d=0nm to d=1nm.

### 5. Conclusion and perspectives

In summary, we reported here an original single-step synthesis and organization process of gold NPs into copolymer thin films. By making vary the polymer solution concentration, it is possible to increase the number of Au NPs into the PS domains and to induce interparticle coupling which leads to the apparition of a second LSPR mode at a wavelength of about 700-750nm. The simulations made with the DDA model showed that the NPs are very close from each other (interparticle distance inferior to 1nm). The ability to load polymer film with 10 nm Au particles into organized domains and to possibly make them interact could make this nanocomposite film appealing for application in non-linear optics and metamaterials. In our study, the copolymer domains long-range order has been lost when filling the copolymer film with gold salts. Efforts should be now made to keep the long-range order of copolymer domains and combine it with the high density of gold colloids, especially for metamaterial fabrication. Another potential for nanocomposite film could be found in plasmonic strain sensor development. Indeed, the use of plasmonic coupling for plasmonic strain sensor has recently attracted leading groups [8, 37]. Indeed, depositing Au NPs onto elastomeric films could provide efficient strain sensors whose color depends on Au NPs interparticle distributions. In our case, the next step will consist in evidencing how sensitive may be the large wavelength peak depending on the inter-NPs distance distributions.

**Acknowledgements**
Financial support of NanoMat (www.nanomat.eu) by the "Ministère de l'enseignement supérieur et de la recherche," the "Conseil régional Champagne-Ardenne," the "Fonds Européen de Développement Régional (FEDER) fund," and the "Conseil général de l'Aube" is acknowledged. The authors also thank the DRRT (Délégation Régionale à la Recherche et à la Technologie) of Champagne-Ardenne for financial support. T. M.

also thanks the CNRS via the chair « optical nanosensors », the UTT-EPF strategic program for supporting the CODEN project and the Labex ACTION project (contract ANR-11-LABX-01-01) for financial support.